\documentclass[12pt]{article}%
\usepackage{amsmath,latexsym}
\usepackage{graphicx}
\usepackage{amsmath}
\usepackage{amsfonts}
\usepackage{amssymb}%
\begin{document}

\title{{Noncommutative geometry and MOND}}
   \author{
Peter K. F. Kuhfittig*\\  \footnote{kuhfitti@msoe.edu}
 \small Department of Mathematics, Milwaukee School of
Engineering,\\
\small Milwaukee, Wisconsin 53202-3109, USA}

\date{}
 \maketitle

\begin{abstract}\noindent
Modified Newtonian dynamics (MOND) is a
hypothesized modification of Newton's law of
universal gravitation to account for the flat
rotation curves in the outer regions of
galaxies, thereby eliminating the need for
dark matter.  Although a highly successful
model, it is not a self-contained physical
theory since it is based entirely on
observations.  It is proposed in this paper
that noncommutative geometry, an offshoot of
string theory, can account for the flat
rotation curves and thereby provide a
justification for MOND.  This paper extends
an earlier heuristic argument by the author.   

\end{abstract}

\section{Introduction}\label{E:introduction}

It is generally assumed that dark matter
in needed to account for the flat galactic
rotation curves in the  outer regions of
galactic halos.  A well-known alternative
is a hypothesis called modified Newtonian
dynamics (MOND) due to M. Milgrom
\cite{mM83}.  Although highly successful,
it is not a self-contained physical
theory, but rather a purely ad hoc variant
based entirely on observations.

The purpose of this paper is to show that
noncommutative geometry, an offshoot of
string theory, can provide a justification
for this theory.  This possibility had
already been suggested in Ref. \cite{pK17}
based on a heuristic argument.  The
dark-matter problem can also be addressed
by means of $f(R)$ and other modified
gravitational theories, even the 
aforementioned noncommutative geometry. 
See, for example, Ref. \cite{cC09} and 
references therein.
%END OF SECTION

\section{Noncommutative geometry and
    the dark-matter hypothesis}
In this section we take a brief look
at noncommutative geometry, as discussed
in Refs. \cite{SS03a, SS03b}.  One
outcome of string theory is that
coordinates may become noncommuting
operators on a $D$-brane \cite{eW96,
SW99}; the commutator is
$[\textbf{x}^{\mu},\textbf{x}^{\nu}]=
i\theta^{\mu\nu}$, where $\theta^{\mu\nu}$
is an antisymmetric matrix.  The main idea,
discussed in Refs. \cite{SS03a, SS03b},
is that noncommutativity replaces
point-like structures by smeared objects,
thereby eliminating the divergences that
normally occur in general relativity.
The smearing effect can be accomplished
in a natural way by means of a Gaussian
distribution of minimal length
$\sqrt{\beta}$ instead of the Dirac delta
function \cite{NSS06, NS10, pK13}.  A
simpler, but equivalent, way is proposed
in Refs. \cite{NM08, pK15}: we assume
that in the neighborhood of the origin,
the energy density of the static
and spherically symmetric and
particle-like gravitational source has
the form
\begin{equation}\label{E:rho1}
  \rho_{\beta}(r)=\frac{m\sqrt{\beta}}
     {\pi^2(r^2+\beta)^2}.
\end{equation}
The point is that the mass $m$ of the
particle is diffused throughout the
region of linear dimension $\sqrt{\beta}$
due to the uncertainty.  According to
Ref. \cite{NSS06}, we can keep the
standard form of the Einstein field
equations in the sense that the Einstein
tensor retains its original form, but
the stress-energy tensor is modified.
It follows that the length scales can
be macroscopic despite the small value
of $\beta$.

As already noted, the gravitational
 source in Eq. (\ref{E:rho1}) results
 in a smeared mass.  According to Refs.
 \cite{SS03a, SS03b}, the Schwarzschild
 solution of the field equations
 involving a smeared source leads to
 the line element
\begin{equation}\label{E:line}
ds^{2}=-\left(1-\frac{2M_{\beta}(r)}{r}\right)dt^{2}
+\frac{dr^2}{1-\frac{2M_{\beta}(r)}{r}}
+r^{2}(d\theta^{2}+\text{sin}^{2}\theta\
d\phi^{2}).
\end{equation}
The smeared mass is given by
\begin{equation}
   M_{\beta}(r)=\int^r_04\pi(r')^2\rho(r')dr'
   =\frac{2M}{\pi}\left(\text{tan}^{-1}
   \frac{r}{\sqrt{\beta}}-
   \frac{r\sqrt{\beta}}{r^2+\beta}\right),
\end{equation}
where $M$ is now the total mass of
the source.  Observe that the mass of
the particle is zero at the center  and
rapidly rises to $M$.  As a result,
from a distance, the smearing is no
longer observed and we get an ordinary
particle: $\text{lim}_{\beta\rightarrow
0}M_{\beta}(r)=M$.

Returning now to the dark-matter
hypothesis, despite its origin in the
1930's, the implications of this
hypothesis were not fully understood
until the 1970's with the discovery
of flat galactic rotation curves, i.e.,
constant velocities sufficiently far
from the galactic center \cite{RTF80}.
This led to the conclusion that matter
in a galaxy increases linearly in the
outward radial direction.  To recall
the reason for this seemingly strange
behavior, suppose $m_1$ is the
mass of a star, $v$ its constant
velocity, and $m_2$ the mass of
everything else in the outer region,
i.e., the region characterized by flat
rotation curves.  Multiplying $m_1$ by
the centripetal acceleration, we get
\begin{equation}
   m_1\frac{v^2}{r}=m_1m_2\frac{G}{r^2},
\end{equation}
where $G$ is Newton's gravitational
constant.  Using geometrized units,
$c=G=1$, we obtain the linear form
\begin{equation}\label{E:flat}
   m_2=rv^2.
\end{equation}
So $v^2$ is independent of $r$, the
distance from the center of the
galaxy.  The purpose of MOND is to
account for this outcome without
hypothesizing dark matter.
%END OF SECTION

\section{Explaining MOND}

Consider next a particle located on
the  spherical surface $r=r_0$.  The
density of the smeared particle now
becomes
\begin{equation}\label{E:rho2}
   \rho_s(r)=\frac{M\sqrt{\beta}}
   {\pi^2[(r-r_0)^2+\beta]^2},
\end{equation}
valid for any surface $r=r_0$.  [If
$r_0=0$, we return to Eq.
(\ref{E:rho1}).]  Eq. (\ref{E:rho2})
can also be interpreted as the density
of the spherical surface, yielding
the smeared mass of the shell in the
outward radial direction, the analogue
of the smeared mass at the origin.
Integration then yields
\begin{equation}\label{E:mass1}
   m_0(r)=\frac{2M}{\pi}
\left[\text{tan}^{-1}
   \frac{r-r_0}{\sqrt{\beta}}-
   \frac{(r-r_0)\sqrt{\beta}}
   {(r-r_0)^2+\beta}\right].
\end{equation}
In particular,
$\text{lim}_{\beta\rightarrow0}m_0(r)
=M$.  So while $m_0(r)$ is zero at
$r=r_0$, it rapidly rises to $M$
in the outward radial direction.
We can readily show that this mass
is completely independent of $r_0$ by
rewriting Eq. (\ref{E:mass1}) as
\begin{equation}\label{E:mass2}
   m_0(r)=\frac{2M}{\pi}
\left[\text{tan}^{-1}
   \frac{1-\frac{r_0}{r}}
   {\frac{\sqrt{\beta}}{r}}-
   \frac{(1-\frac{r_0}{r})
   \frac{\sqrt{\beta}}{r}}
   {(1-\frac{r_0}{r})^2+
   (\frac{\sqrt{\beta}}{r})^2}
   \right].
\end{equation}
Indeed, as $\frac{\sqrt{\beta}}{r}
\rightarrow 0$, $m_0(r)\rightarrow
M$ for every $r_0$.  To retain the
smearing, we would normally require
that $\beta >0$.  So for $m_0(r)$
to get close to $M$, $r=r_0$ would
have to be sufficiently large.  In
other words, we would have to be in
the outer region of the galaxy, i.e.,
the region characterized by flat
rotation curves.

Now consider the finite sequence
$\{r_i\}$ of such radii.  Then the
smeared mass of every spherical
shell becomes
\begin{equation}\label{E:mass2}
   m_i(r)=\frac{2M}{\pi}
\left[\text{tan}^{-1}
   \frac{r-r_i}{\sqrt{\beta}}-
   \frac{(r-r_i)\sqrt{\beta}}
   {(r-r_i)^2+\beta}\right]
\end{equation}
with $\text{lim}_{\beta\rightarrow0}
m_i(r)=M$ for every $i$.  To obtain
the total mass $M_T$ of the outer
region, we can think of $m_i(r)$ as
the increase in $M_T$ per unit length
in the outward radial direction,
making $M$ a dimensionless constant.
If we denote the thickness of each
smeared spherical shell by
$\Delta r$, then $m_i(r)\Delta r$
becomes the mass of the shell.
However, we cannot simply integrate
$m_i(r)$ over the entire region,
since each shell has a different
$r_i$.  Instead, we proceed as
follows:
\begin{multline}
   \Delta M_T=\int_{r_i}^{r_i+\Delta r}
   m_i(r)\,dr
   =\int_{r_i}^{r_i+\Delta r}
   \frac{2M}{\pi}
   \left[\text{tan}^{-1}
   \frac{r-r_i}{\sqrt{\beta}}-
   \frac{(r-r_i)\sqrt{\beta}}
   {(r-r_i)^2+\beta}\right]dr\\
   =\frac{2M}{\pi}\left[(r-r_i)
   \text{tan}^{-1}\frac{r-r_i}
   {\sqrt{\beta}}\left.-\sqrt{\beta}\,\,
   \text{ln}\left[(r-r_i)^2+
   \beta\right]
   \right]\right|_{r_i}^{r_i+\Delta r} \\
   = \frac{2M}{\pi}\Delta r\left[
   \text{tan}^{-1}
   \frac{\Delta r}{\sqrt{\beta}}
   -\sqrt{\beta}\,\frac{\text{ln}
   \left[(\Delta r)^2+\beta\right]}
   {\Delta r}+\frac{\sqrt{\beta}\,
   \text{ln}\,\beta}{\Delta r}\right].
\end{multline}
Given that $\Delta r$ is large compared
to $\beta$ and that
$\text{lim}_{\beta\rightarrow0}
\sqrt{\beta}\,\,\text{ln}\,\beta=0$,
it follows that \begin{equation}
   \Delta M_T\approx M\Delta r
\end{equation}
for all $r$ in the outer region.  So
in this region, all shells of thickness
$\Delta r$ have approximately the same
mass $\Delta M_T$ due to the
noncommutative-geometry background.
Since $\beta>0$ is fixed, we can now
safely let $\Delta r\rightarrow 0$.
This leads to our main conclusion:
from  $\text{lim}_{\Delta r\rightarrow0}
\Delta M_T/\Delta r=M$, we obtain the
linear form
\begin{equation}
   M_T=Mr.
\end{equation}
By Eq. (\ref{E:flat}), $M=v^2$, showing
that $v^2$ is indeed independent of $r$,
in agreement with MOND in the outer region 
of the galaxy.  This region is characterized 
by extremely low accelerations, also referred 
to as the deep-MOND regime.
%END OF SECTION

\section{Conclusions}

The existence of flat rotation curves
in the outer regions of galaxies can
be accounted for by the presence of
dark matter or by the use of modified
gravitational theories.  An example of
the latter is M. Milgrom's modified
Newtonian dynamics or MOND.  Although \
a highly successful model, MOND is a
purely ad hoc theory based entirely on
observations.  So it cannot be called
a self-contained theory.

It is proposed in this paper that
noncommutative geometry, an offshoot
of string theory, can account for the
flat rotation curves.  Viewed as a 
modification of Einstein's theory, 
noncommutative geometry could 
therefore provide an explanation 
for MOND.

\end{document}